\documentclass[12pt,a4paper]{article}
\pdfoutput=1
\usepackage[utf8]{inputenc}
\usepackage{float}
\usepackage{amsmath,graphicx,dcolumn,subcaption}
\usepackage{siunitx}
\usepackage[mathscr]{eucal}
\usepackage{cite}
\usepackage{color}
\usepackage{rotating}
\usepackage{ulem}
\usepackage[numbers,sort&compress]{natbib}
\usepackage{setspace}
\newcommand\as{\alpha_{\mathrm{S}}} 
\newcommand\f[2]{\frac{#1}{#2}} 
 
\def\to{\rightarrow} 
\def\nn{\nonumber}

\def\ito{\leftarrow}
\def\rcut{r_{\rm cut}}

\def\zmin{z_{\rm min}}
\def\zmax{z_{\rm max}}
\def\sigmah{{\hat \sigma}}

\usepackage{hyperref}
\usepackage{footnotebackref}

\begin{document} 
\begin{titlepage}
\begin{flushright}
ZU-TH 47/19\\
\end{flushright}

\renewcommand{\thefootnote}{\fnsymbol{footnote}}
\vspace*{2cm}

\begin{center}
  {\Large \bf The $q_T$ subtraction method: electroweak corrections\\[0.3cm] and power suppressed contributions
  }
\end{center}

\par \vspace{2mm}
\begin{center}
  {\bf
    Luca Buonocore${}^{(a,b)}$},
  {\bf Massimiliano Grazzini${}^{(b)}$}
  and  
  {\bf Francesco Tramontano${}^{(a)}$
  }

\vspace{5mm}

${}^{(a)}$Dipartimento di Fisica, Universit\`a di Napoli Federico II and\\INFN, Sezione di Napoli, I-80126 Napoli, Italy

${}^{(b)}$Physik Institut, Universit\"at Z\"urich, CH-8057 Z\"urich, Switzerland

\vspace{5mm}

\end{center}

\par \vspace{2mm}
\begin{center} {\large \bf Abstract}

\end{center}
\begin{quote}
\pretolerance 10000

Building upon the formulation of transverse-momentum resummation for heavy-quark hadroproduction, we present the first application of the $q_T$ subtraction formalism to the computation of electroweak corrections to massive lepton pairs through the Drell-Yan mechanism. We then study the power suppressed contributions to the $q_T$ subtraction formula in the parameter $\rcut$, defined as the minimum transverse momentum of the lepton pair normalised to its invariant mass.
We analytically compute the leading power correction from initial and final-state radiation to the inclusive cross section. In the case of initial-state radiation the power correction is {\it quadratic} in $\rcut$ and our analytic result is consistent with results previously obtained in the literature.
Final-state radiation produces {\it linear} contributions in $\rcut$ that may challenge the efficiency of the $q_T$ subtraction procedure.
We explicitly compute the linear power correction in the case of the inclusive cross section and we discuss the extension
of our calculation to differential distributions.

\end{quote}

\vspace*{\fill}
\begin{flushleft}
November 2019
\end{flushleft}
\end{titlepage}

\section{Introduction}
\label{sec:intro}

The $q_T$ subtraction formalism~\cite{Catani:2007vq} is
a method to handle and cancel the IR divergences appearing in QCD computations at next-to-next-to-leading order (NNLO) and beyond.
In its original formulation it has been applied to carry out a variety of NNLO QCD computations for the production of colourless final states in hadronic collisions \cite{Catani:2009sm,Catani:2011qz,Catani:2018krb,Ferrera:2011bk,Grazzini:2013bna,Ferrera:2014lca,Grazzini:2015nwa,Cascioli:2014yka,Gehrmann:2014fva,Grazzini:2015hta,Grazzini:2016ctr,Grazzini:2016swo,Grazzini:2017ckn,Kallweit:2018nyv,deFlorian:2016uhr}.
Most of the  above computations are now publicly available in {\sc Matrix} \cite{Grazzini:2017mhc}.
A first application of $q_T$ subtraction to the computation of the approximate next-to-next-to-next-to-leading order (N$^3$LO) QCD corrections to Higgs boson production through gluon fusion has been presented recently \cite{Cieri:2018oms}.

In the last few years, thanks to the formulation of transverse-momentum resummation for heavy-quark production \cite{Zhu:2012ts,Li:2013mia,Catani:2014qha,Angeles-Martinez:2018mqh,inprep}
the method has been extended and applied to the production of top-quark pairs \cite{Bonciani:2015sha,Catani:2019iny,Catani:2019hip}.
The $q_T$ subtraction counterterm is constructed by exploiting the universal behavior of the associated transverse-momentum ($q_T$) distribution. Therefore, the subtraction is intrinsically {\it non local}
and in practice the computation is carried out by introducing a cut, $\rcut$ on the tranverse momentum of the colourless system normalised to its invariant mass.
When evaluated at finite $\rcut$ both the contribution of the real emission and the one of the counterterm exibit logarithmically divergent terms plus additional power suppressed contributions that vanish as $\rcut\to 0$.
The efficiency of the subtraction procedure crucially depends on the size of such power suppressed contributions.

In the inclusive production of a colourless final state the power suppressed contributions are known to be {\it quadratic} in $\rcut$ (modulo logarithmic enhancements) \cite{Grazzini:2016ctr}.
This allows us to obtain precise predictions by either evaluating the cross section at sufficiently small $\rcut$, or carrying out the $\rcut\to 0$ extrapolation \cite{Grazzini:2017mhc}
\footnote{The only exception is the production of direct photons ($\gamma\gamma$ \cite{Catani:2011qz,Catani:2018krb}, $Z\gamma$ \cite{Grazzini:2013bna}, $W\gamma$ \cite{Grazzini:2015nwa}....), for which a fully inclusive cross section cannot be defined, and an isolation prescription is required. The interplay of the isolation prescription with the subtraction procedure makes the $\rcut$ dependence stronger \cite{Grazzini:2017mhc,Ebert:2019zkb}.}.
The power suppressed contributions to the next-to-leading order (NLO) total cross section have been explicitly evaluated in Refs.~\cite{Ebert:2018gsn,Cieri:2019tfv}.
In the case of heavy-quark production the $\rcut$ dependence is found to be {\it linear} \cite{Catani:2017tuc,Catani:2019iny,Catani:2019hip}, and it is an interesting question
to investigate the origin of this peculiar behavior.

Up to now the $q_T$ subtraction formalism has been applied only to higher-order QCD computations.
The formulation of the method for heavy-quark production can be straightforwardly extended to the computation of NLO electroweak (EW) corrections to the Drell-Yan process.
The purpose of the present paper is twofold. We first present and discuss the first application of the $q_T$ subtraction formalism to
the computation of the NLO EW corrections to the production of massive lepton pairs. Our computation consistently
includes initial-state radiation, final-state radiation from the massive leptons and their interference, and our results
are compared to an independent computation that we carry out with the dipole subtraction formalism \cite{Catani:2002hc}.
Then, we present the analytic computation of the power suppressed contributions, and we confirm the linear $\rcut$ behaviour by computing its NLO coefficient.
We also extend our results to the case in which cuts are applied.

The paper is organised as follows. In Sec.~\ref{sec:qt} we review the $q_T$-subtraction formalism, by detailing its implementation up to NLO in the case of heavy-quark production.
In Sec.~\ref{sec:nloew} we apply the formalism to the computation of NLO EW corrections to the Drell-Yan process. In Sec.~\ref{sec:power} we study the power suppressed contributions and we explitly compute the leading
power corrections in the case of final-state and initial-state radiation. In Sec.~\ref{sec:summa} we summarise our results.

\section{The $q_T$ subtraction formalism}
\label{sec:qt}

The $q_T$ subtraction formalism~\cite{Catani:2007vq} is
a method to handle and cancel the IR divergences appearing in higher-order QCD computations.
The method uses IR subtraction counterterms that are constructed by considering
and explicitly computing the transverse-momentum $q_T$ distribution
of the produced final-state system.
At Born level such distribution is proportional to $\delta(q_T^2)$.
At higher perturbative orders multiple radiation of soft and collinear partons makes
the distribution divergent in the $q_T \rightarrow 0$ limit.
If the produced final-state system is composed of non-QCD (colourless) partons
(e.g., leptons, vector bosons or Higgs bosons), the small-$q_T$ behaviour has a universal 
(process-independent) structure that is explicitly known up to the NNLO level (and, in part, at N$^3$LO \cite{Cieri:2018oms,Billis:2019vxg})
through the formalism of transverse-momentum resummation 
\cite{Catani:2013tia}. These results on transverse-momentum resummation
are sufficient to fully specify the $q_T$ subtraction  formalism for this entire
class of processes.
By using the formulation of transverse-momentum resummation
for heavy-quark production \cite{Zhu:2012ts,Li:2013mia,Catani:2014qha,Angeles-Martinez:2018mqh,inprep},
the $q_T$ subtraction formalism has been recently extended to this class of processes \cite{Bonciani:2015sha,Catani:2019iny,Catani:2019hip}.

According to the $q_T$ subtraction method~\cite{Catani:2007vq}, the parton level
differential cross section $d{\sigma}^{Q{\bar Q}}_{(N)NLO}$ for 
the inclusive production 
process $pp\to Q{\bar Q}+X$
can be written as
\begin{equation}
\label{eq:main}
d{\sigmah}^{Q{\bar Q}}_{(N)NLO}={\cal H}^{Q{\bar Q}}_{(N)NLO}\otimes d{\sigmah}^{Q{\bar Q}}_{LO}
+\left[ d{\sigmah}^{Q{\bar Q}+\rm{jet}}_{(N)LO}-
d{\sigmah}^{Q{\bar Q}, \, CT}_{(N)NLO}\right],
\end{equation}
where $d{\sigmah}^{Q{\bar Q}+\rm{jet}}_{(N)LO}$ is the $Q{\bar Q}$+jet cross 
section 
at (N)LO accuracy.
The square bracket term of Eq.~(\ref{eq:main}) is IR finite in the limit
$q_T \to 0$, but its individual contributions,
$d{\sigmah}^{Q{\bar Q}+\rm{jet}}_{(N)LO}$ and
$d{\sigmah}^{Q{\bar Q}, \, CT}_{(N)NLO}$, are separately divergent.
The IR subtraction counterterm $d{\sigmah}^{Q{\bar Q}, \,CT}_{(N)NLO}$
is obtained from the (N)NLO perturbative expansion 
(see, e.g., Refs.~\cite{Bozzi:2005wk,Bonciani:2015sha})
of the resummation formula
of the logarithmically-enhanced
contributions to the $q_T$ distribution
of the $Q{\bar Q}$ pair \cite{Zhu:2012ts,Li:2013mia,Catani:2014qha}:
the explicit form of $d{\sigmah}^{Q{\bar Q}, \,CT}_{(N)NLO}$ 
can be completely worked out up to NNLO accuracy.

In the following we will limit ourselves to consider Eq.~(\ref{eq:main}) up to NLO accuracy.
The explicit expression of $d{\sigmah}^{Q{\bar Q}, \,CT}_{NLO}$ in the partonic channel $ab \to Q {\bar Q}+X$ reads \cite{Bonciani:2015sha}
\begin{equation}
\label{eq:ctnlo}
d{\sigmah}^{Q{\bar Q}, \,CT}_{NLO\, ab}=\sum_{c=q,{\bar q},g}
\f{\as}{\pi}\;
\Sigma^{(1)}_{c{\bar c}\ito ab} \otimes d{\sigmah}^{Q{\bar Q}}_{LO\, c{\bar c}}\; \f{dq_T^2}{M^2}\;\;,
\end{equation}
where $M$ is the invariant mass of the $Q {\bar Q}$ pair and the symbol $\otimes$ denotes convolutions with respect to the longitudinal-momentum fractions $z_1$ and $z_2$ of the colliding partons.
The functions $\Sigma^{(1)}_{c{\bar c}\ito ab}$ in Eq.~(\ref{eq:ctnlo}) can be written as
\begin{equation}
  \label{eq:sigma1}
  \Sigma^{(1)}_{c{\bar c}\ito ab}(z_1,z_2;r)=\Sigma^{(1,2)}_{c{\bar c}\ito ab}(z_1,z_2) {\tilde I}_2(r)+\Sigma^{(1,1)}_{c{\bar c}\ito ab}(z_1,z_2){\tilde I}_1(r)
\end{equation}
where $r=q_T/M$, and the coefficients $\Sigma^{(1,k)}_{c{\bar c}\ito ab}(z_1,z_2)$ ($k=1,2$) read
\begin{equation}
  \Sigma^{(1,2)}_{c{\bar c}\ito ab}(z_1,z_2)=-\f{1}{2}A_c^{(1)}\delta_{ca}\delta_{{\bar c}b}\delta(1-z_1)\delta(1-z_2)\\
  \label{eq:sig12}
\end{equation}
\begin{align}
\Sigma^{(1,1)}_{c{\bar c}\ito ab}(z_1,z_2)=&-\Big[\delta_{ca}\delta_{{\bar c}b}\delta(1-z_1)\delta(1-z_2)\, B^{(1)}_c+\delta_{ca}\delta(1-z_1)\,P_{{\bar c}b}^{(1)}(z_2)+\delta_{{\bar
      c}b}\delta(1-z_2)\,P_{ca}^{(1)}(z_1)\Big]\nn\\
&-\delta_{ca}\delta_{{\bar c}b} \delta(1-z_1)\delta(1-z_2)
\frac{\langle {\cal M}_{c{\bar c}\to Q{\bar Q}}|\left({\bf \Gamma}^{(1)}_t+{\bf \Gamma}_t^{(1)\dagger}\right)|{\cal M}_{c{\bar c}\to Q{\bar Q}}\rangle}{|{\cal M}_{c{\bar c}\to Q{\bar Q}}|^2}\, .
  \label{eq:sig11}
\end{align}
The coefficients
$A_c^{(1)}$ and $B_c^{(1)}$ are the first-order coefficients for transverse-momentum resummation ($A^{(1)}_q=C_F$, $A^{(1)}_g=C_A$, $B^{(1)}_q=-3/2\, C_F$, $B^{(1)}_g=-(11/6\, C_A-n_F/3)$).
The functions $P_{ab}^{(1)}(z)$ are the lowest-order DGLAP kernels 
(the overall normalisation is specified according to the notation in Eq.~(41) of
Ref.~\cite{Bozzi:2005wk}).
The functions ${\tilde I}_k(r)$ $(k=1,2)$, which appear in 
Eq.~(\ref{eq:sigma1}),
encapsulate the singular behavior at small $q_T$, and they read (see Appendix~B of Ref.~\cite{Bozzi:2005wk})
\begin{equation}
  \tilde{I}_1(r) = -\frac{b_0}{r}K_1(b_0 r), \quad  \tilde{I}_2(r) = \frac{2b_0}{r}\left[K_1(b_0 r)\ln{r}-
    \frac{\partial K_\nu(b_0 r)}{\partial \nu}\bigg|_{\nu=1}\right],
\end{equation}
where $b_0= 2e^{-\gamma_E}$ and we have introduced the modified Bessel function of imaginary argument
\begin{equation}
    K_\nu(r) = \int_0^\infty dt e^{-r\cosh{t}}\cosh{\nu t}\, .
\end{equation}
The coefficient $\Sigma^{(1,2)}_{c{\bar c}\ito ab}(z_1,z_2)$ in Eq.~(\ref{eq:sig12}) controls the leading logarithmic contribution at small $q_T$, while the coefficient
$\Sigma^{(1,1)}_{c{\bar c}\ito ab}(z_1,z_2)$ in Eq.~(\ref{eq:sig11}) controls the next-to-leading logarithmic term. The latter has a first term (first line in Eq.~(\ref{eq:sig11}))
which is identical to what we have in the case of the production of a colour singlet. The second term (second line in Eq.~(\ref{eq:sig11})) is due to soft radiation and it
is an additional term that is specific of the $q_T$ subtraction method for the
case of heavy-quark pair production \cite{Bonciani:2015sha}.
Here ${\bf \Gamma}^{(1)}_t$
is the first-order contribution to the soft anomalous dimension for transverse-momentum 
resummation in heavy-quark production and its explicit expression 
is given in Eq.~(33) of Ref.~\cite{Catani:2014qha}.
The soft anomalous dimension is a matrix acting on the colour indeces of the four hard partons in the Born level scattering amplitude
$|{\cal M}_{c{\bar c}\to Q{\bar Q}}\rangle$.
At this perturbative order the soft anomalous dimension is expressed in terms of colour correlators ${\bf T}_i\cdot {\bf T}_j$ with definite kinematic dependence,
where the indices $i$ and $j$ refer to the hard-scattering partons.

The first-order
hard-collinear coefficients
${\cal H}^{Q{\bar Q}}_{NLO}$ in Eq.~(\ref{eq:main}) are also completely known 
\cite{Zhu:2012ts,Li:2013mia,Catani:2014qha}. In the next Section we apply the method to the computation of EW corrections to the Drell-Yan process.

\section{NLO EW corrections to the Drell-Yan process}
\label{sec:nloew}

We consider the hadroproduction of a dilepton pair through the Drell-Yan mechanism.
NLO EW corrections to this process have been considered in Refs.~\cite{Baur:2001ze,Zykunov:2005tc,CarloniCalame:2007cd,Arbuzov:2007db}.
A tuned comparison of various Monte Carlo codes is presented in Ref.~\cite{Alioli:2016fum}.

The NLO $q_T$ subtraction formalism for heavy-quark production reviewed in Sect.~\ref{sec:qt}
can be straightforwardly extended to the computation of the NLO EW corrections to the Drell-Yan process.
In this case the heavy-quark pair is replaced by a massive lepton pair and the abelian limit is carried out along the lines of Ref.~\cite{deFlorian:2018wcj}.
The partonic cross section up to NLO EW can be evaluated by using
\begin{equation}
\label{eq:mainew}
d{\sigmah}_{NLO}={\cal H}_{NLO}\otimes d{\sigmah}_{LO}
+\left[ d{\sigmah}^{R}-
d{\sigmah}^{CT}\right],
\end{equation}
where $d\sigmah^{R}$ is the real emission cross section and the functions ${\cal H}_{NLO}$ and $d{\sigmah}^{CT}$ are
obtained from the corresponding functions appearing in Eq.~(\ref{eq:main}) with the replacements
\begin{equation}
  C_A\to 0~~~~~~~~~~C_F\to e_f^2~~~~~~~~~~{\bf T}_i^2\to e_i^2~~~~~~~~~~{\bf T}_i\cdot {\bf T}_j\to e_i e_j\, .
\end{equation}
As is well known, at LO (i.e. ${\cal O}(\alpha^2)$) both the $q{\bar q}$ and the $\gamma\gamma$ partonic channels contribute and we can write for the hadronic cross section
\begin{equation}
\sigma_{LO}=\sigma_{LO}^{q{\bar q}}+\sigma^{\gamma\gamma}_{LO}\, ,
\end{equation}
where $\sigma_{LO}^{q{\bar q}}$ and $\sigma^{\gamma\gamma}_{LO}$ are the Born level cross sections in the $q{\bar q}$ and $\gamma\gamma$ channels, respectively.
At NLO EW we can write
\begin{equation}
  \sigma_{NLO}=\sigma_{LO}^{q{\bar q}}+\sigma^{\gamma\gamma}_{LO}+\Delta\sigma_{q{\bar q}}+\Delta\sigma_{q\gamma}+\Delta\sigma_{\gamma\gamma}
\end{equation}
where we have introduced the ${\cal O}(\alpha^3)$ correction in the $q{\bar q}$ channel, $\Delta\sigma_{q{\bar q}}$, the corresponding correction in the $q({\bar q})\gamma$ channel, $\Delta\sigma_{q\gamma}$,
and the correction in the $\gamma\gamma$ channel, $\Delta\sigma_{\gamma\gamma}$. Since the $\gamma\gamma$ channel provides only a very small contribution to the Drell-Yan cross section, $\Delta\sigma_{\gamma\gamma}$
will be neglected in the following discussion.

Our calculation is carried out by using an extension of the numerical program of Ref.~\cite{Catani:2009sm}.
All the required tree level matrix elements are computed analytically while the virtual EW corrections for $q{\bar q}\to  l^+ l^-$, which include vertex and box diagrams, are obtained by using {\sc Gosam} \cite{Cullen:2011ac,Cullen:2014yla}.
We use the setup of Ref.~\cite{Dittmaier:2009cr}, and, in particular, we work in the $G_\mu$ scheme 
with
\begin{align}
  & G_F=1.16637\times 10^{-5}~{\rm GeV}^{-2}  & \alpha(0)=1/137.03599911\\
  & m_W=80.403~{\rm GeV}                     & m_Z=91.1876~{\rm GeV}\\
  & \Gamma_W=2.141~{\rm GeV}                 & \Gamma_Z=2.4952~{\rm GeV}
\end{align}
and use the complex-mass scheme~\cite{Denner:2005fg} throughout. More precisely, real and virtual photons emissions are controlled by $\alpha(0)$, while the $\alpha^2$ in the LO cross section is derived from $G_F$, $m_Z$ and $m_W$.
Following Ref.~\cite{Dittmaier:2009cr}, the MRST2004qed \cite{Martin:2004dh} parton distribution functions (PDFs) are used.
In order to avoid large logarithmic terms in the lepton mass which may complicate the numerical convergence
we set the mass of the final-state lepton to $m_l=10$ GeV.
We also use bare leptons (i.e. no recombination with the photon is used).

\renewcommand{\arraystretch}{1.6}
\begin{table}[H]
  \centering
\begin{tabular}{|c|c|c|}
  \hline
  & $q_T+\text{GoSam}$  &  {\sc cs}+{\sc Recola}\\
  \hline
  $\sigma_{LO}^{q\bar{q}}$ (pb) & 
   \multicolumn{2}{c|}{$683.53 \pm 0.03$} \\
  \hline
  $\Delta\sigma_{q\overline{q}}$ (pb) & $-5.920 \pm 0.034$ &  $-5.919 \pm 0.008$ \\
  \hline
  $\sigma_{LO}^{\gamma\gamma}$ (pb) & 
   \multicolumn{2}{c|}{$ 1.1524 \pm 0.0004 $} \\
  \hline
  $\Delta\sigma_{q\gamma}$ (pb) & $-0.6694 \pm 0.0008$ & $-0.6690 \pm 0.0005$ \\
  \hline
\end{tabular}
\caption{\label{tab:cmp_deltaqq} \small Comparison of NLO EW corrections to the Drell-Yan process computed with $q_T$ subtraction and dipole subtraction. In the $q{\bar q}$ channel the $q_T$ result is obtained with a linear extrapolation in the $\rcut\to 0$ limit (see Figure~\ref{figs:rcut-var-EW}),
while in the $q({\bar q})\gamma$ channel it is obtained at $\rcut = 0.01 \%$. The LO result in the $q{\bar q}$ and $\gamma\gamma$ channels is also reported for reference.}
\end{table}
\noindent The following set of cuts are applied
\begin{equation}\label{eq:cuts}
  m_{ll}>50~{\rm GeV}~~~~~~~~~~p_{T,l}>25~{\rm GeV}~~~~~~~~~~~~|y_l|<2.5\, .
\end{equation}
To validate our implementation, we have repeated our calculation by using the dipole subtraction method \cite{Catani:2002hc} and the independent matrix-element generator {\sc Recola}~\cite{Actis:2012qn,Actis:2016mpe} for the virtual corrections.
In Table~\ref{tab:cmp_deltaqq} we report our result for the lowest order cross sections $\sigma^{q{\bar q}}_{LO}$ and $\sigma^{\gamma\gamma}_{LO}$, and the NLO EW corrections in the $q{\bar q}$ and $q\gamma$ channels, $\Delta\sigma_{q\overline{q}}$ and $\Delta\sigma_{q\gamma}$. 
The NLO correction $\Delta\sigma_{q{\bar q}}$ is obtained performing the calculation at different values of $\rcut$ and extrapolating to $\rcut\to 0$ through a linear fit.
Our results are compared with the corresponding results obtained with dipole subtraction (CS+{\sc Recola}).
We see that the two results are in perfect agreement.
Furthermore, to have a fully independent validation, we have repeated our calculation by setting the mass of the heavy lepton
to the physical muon mass  $m_l\equiv m_\mu=105.658369\,$MeV, and we 
compared our results with those obtained with the well established 
public generator {\sc Sanc} \cite{Andonov:2004hi}. We use the setup and fiducial 
cuts as before. Our results are reported in Table~\ref{tab:cmp_sanc}, and show that the agreement is quite good, at few {\it per mille} on the NLO correction.

\renewcommand{\arraystretch}{1.6}
\begin{table}[H]
\centering
\begin{tabular}{|c|c|c|}
  \hline
  & $q_T+\text{GoSam}$  & {\sc Sanc}\\
  \hline
  $\Delta\sigma_{q\overline{q}}+\Delta\sigma_{q\gamma}$ (pb) & $-29.95 \pm 0.04$ &  $-29.99 \pm 0.02$ \\
  \hline
\end{tabular}
\caption{\label{tab:cmp_sanc} \small Tuned comparison for NLO EW corrections to the Drell-Yan process with $m_l = m_\mu=105.658369\,$ MeV with the {\sc Sanc} generator. The $q_T$ result is the limiting value for $\rcut\to 0$ obtained with a linear fit for the NLO correction in the diagonal $q\bar{q}$-annihilation channel, and it is the value at $\rcut = 0.01 \%$ for the off-diagonal $q({\bar q})\gamma$ channel.}
\end{table}

\begin{figure}[H]
  \centering
  \includegraphics[width=0.48\textwidth]{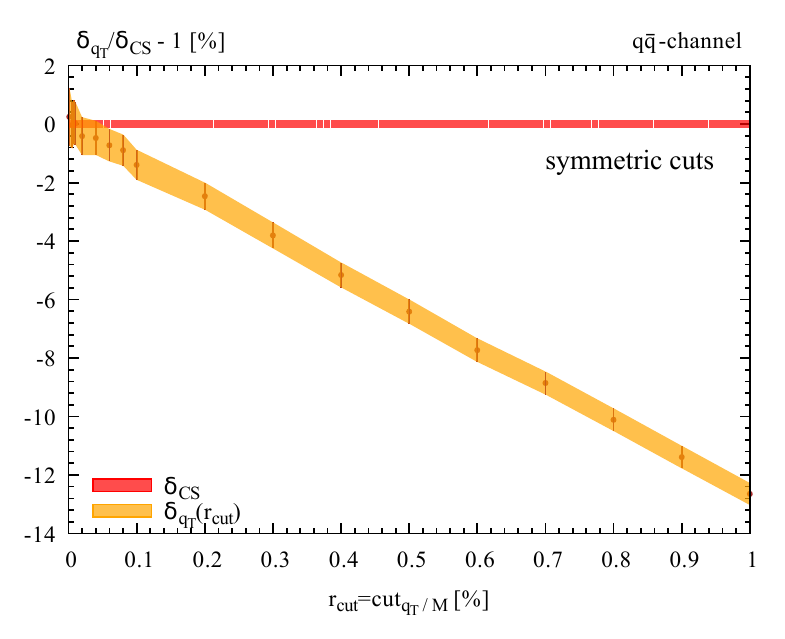}
  \includegraphics[width=0.48\textwidth]{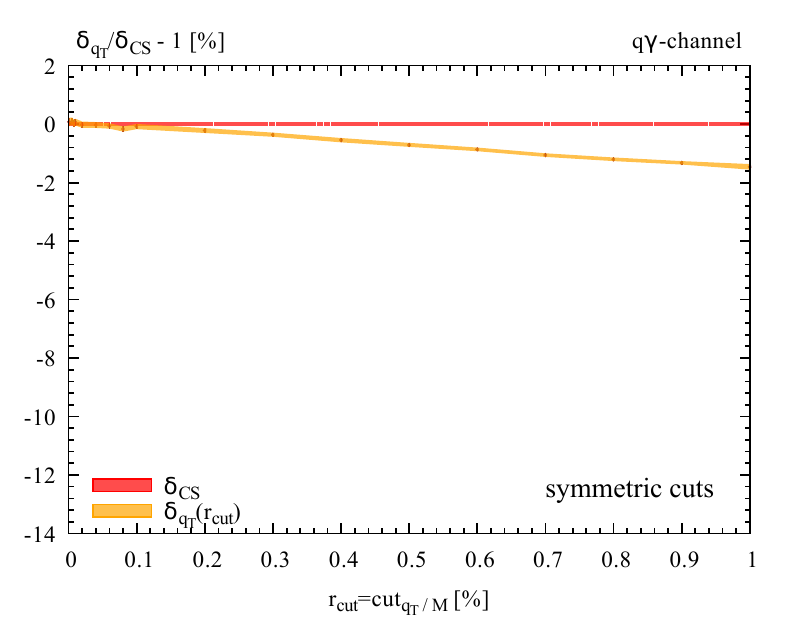}
  \caption{\label{figs:rcut-var-EW} \small NLO EW correction as a function of $\rcut$ in the dominant $q\overline{q}$ diagonal channel (left panel) and in the off-diagonal $q({\bar q})\gamma$ channel (right panel) at $14\,$TeV. The NLO result is normalised to the $\rcut$-independent cross section computed with dipole subtraction. The lepton mass is fixed to $m_l=10\,$GeV. The fiducial cuts in Eq.~\eqref{eq:cuts} are applied.} 
\end{figure}

\begin{figure}
  \centering
  \includegraphics[width=0.48\textwidth]{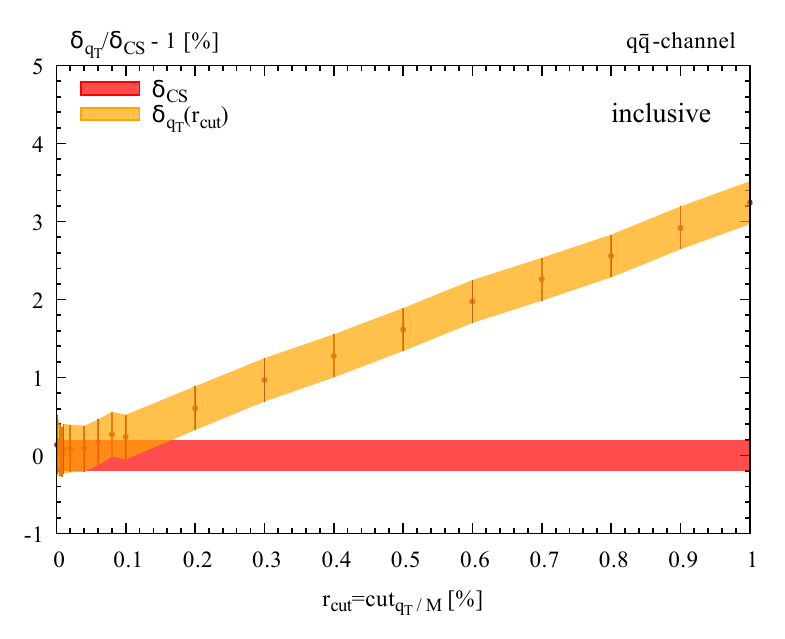}
  \includegraphics[width=0.48\textwidth]{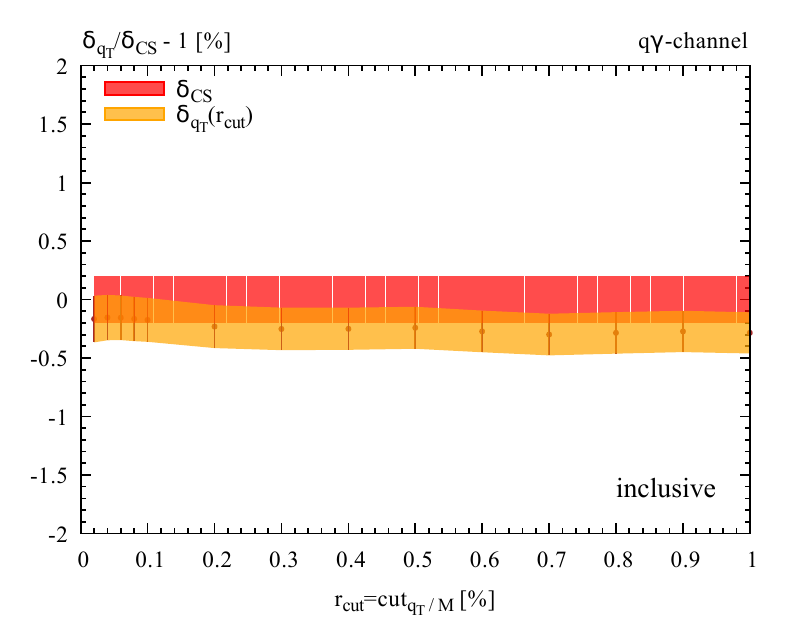}
  \caption{\label{figs:rcut-var-EW-nocuts} \small NLO EW correction as a function of $\rcut$ in the dominant $q\overline{q}$ diagonal channel (left panel) and in the off-diagonal $q({\bar q})\gamma$ channel (right panel) at $14\,$TeV. The NLO result is normalised to the $\rcut$-independent cross section computed with dipole subtraction. The lepton mass is fixed to $m_l=10\,$GeV. No cuts are applied.}
\end{figure}

We have studied the dependence of the NLO corrections for the fiducial cross section on $\rcut$.
We have varied $\rcut$ in the range $0.01\% \leq \rcut \leq 1\%$ and we have used the $\rcut$-independent cross section computed
with our inhouse implementation of the dipole subtraction method as normalisation.
In order to avoid complications due to the small lepton mass, which could obscure the $\rcut$ behavior, here and in the following we stick to $m_l=10$ GeV.
The results for the $\rcut$ dependent correction $\delta_{q_T}=\Delta\sigma/\sigma_{LO}^{q{\bar q}}$ in the $q{\bar q}$ and $q\gamma$ channels
are shown in Figure~\ref{figs:rcut-var-EW}. A distictive linear behavior in the dominant 
$q\bar{q}$-annihilation channel emerges. Nonetheless, as reported in Ref.~\cite{Grazzini:2017mhc}, 
it is known that symmetric cuts on the transverse momenta of the final 
state leptons challenge the convergence of $q_T$-subtraction leading 
to a stronger dependence on $\rcut$ even in the case in which a charge-neutral final state is produced.
In Figure~\ref{figs:rcut-var-EW-nocuts} we show the dependence of the NLO 
corrections for the inclusive cross section on $\rcut$ when no cuts 
are applied. Again a distinct linear behavior in the dominant 
$q\bar{q}$-annihilation channel emerges, in agreement with what has already 
been observed for the case of the $t{\bar t}$ cross section \cite{Catani:2019iny}, which 
can be clearly interpreted as a genuine new effect due to the 
emission of radiation off the massive final-state leptons.

\section{Power corrections}
\label{sec:power}

In this Section we analytically study the behavior of NLO cross sections computed with $q_T$ subtraction in the $\rcut\to 0$ limit.
We are interested in determining the structure of the leading power correction to the inclusive cross section, and to identify the origin of the linear behavior observed in Sec.~\ref{sec:nloew}.

We recall that when applying the $q_T$ subtraction formula the second term on the right hand side of Eq.~(\ref{eq:mainew}) is computed by introducing a lower limit $\rcut$ on the $q_T/M$ ratio.
With such a cutoff we can treat separately the real contribution $d{\hat \sigma}^R$ and the counterterm $d{\hat \sigma}^{CT}$.
We start our discussion from the contribution of the counterterm. From Eq.~(\ref{eq:ctnlo}) we have
\begin{equation}
  d{\hat \sigma}_{ab}^{CT}(\rcut)=\sum_{c=q,{\bar q},\gamma}\int_{\rcut}^\infty 2rdr\, \f{\as}{\pi} \Sigma^{(1)}_{c{\bar c}\ito ab} \otimes d{\hat \sigma}^{l^+l^-}_{LO\, c{\bar c}}\, .
\end{equation}
The NLO coefficient $\Sigma^{(1)}_{c{\bar c}\ito ab}$ depends on $r=q_T/M$ only through the functions ${\tilde I}_i(r)$. Therefore we have
\begin{equation}
  \f{d{\hat \sigma}_{ab}^{CT}(\rcut)}{d\rcut}=-2\rcut \f{\as}{\pi}\left(\Sigma^{(1,2)}_{c{\bar c}\ito ab}{\tilde I}_2(\rcut)+ \Sigma^{(1,1)}_{c{\bar c}\ito ab}{\tilde I}_1(\rcut)\right)\otimes d{\hat \sigma}^{l^+l^-}_{LO\, c{\bar c}}\,.
\end{equation}
In the small $r$ limit the integrals ${\tilde I}_1(r)$ and ${\tilde I}_2(r)$ read
\begin{align}
\label{eq:I1I2ap}
     \tilde{I}_1(r) &= -\frac{1}{r^2} + \frac{b_0^2}{4}\left(1-2\ln{r}\right) + O(r^2),\nn\\
     \tilde{I}_2(r) &= \frac{4\ln{r}}{r^2} + \frac{b_0^2}{2}\left(-1+2\ln^2{r}\right) + O(r^2)\, ,
\end{align}
i.e., they depend quadratically on $r$ modulo logarithmic terms. This results holds also at NNLO and beyond.
It follows that the leading power corrections from the counterterm are always quadratic in $\rcut$, independently on the perturbative order.
As a consequence, the linear behavior with $\rcut$ that we observe in heavy-quark production and in the EW corrections to dilepton production must be due to the real emission. 
In the following we analytically compute the real-emission contribution at small values of $\rcut$.

We consider the production of a massive lepton pair in pure QED in the diagonal channel
  \begin{equation}
    q(p_1)+{\bar q}(p_2)\to l^+(p_3) l^-(p_4)+\gamma(k)
    \end{equation}
  with $p_3^2=p_4^2=m^2$. We define the variables
  \begin{equation}
    s=(p_1+p_2)^2~~~~~~~M^2=(p_3+p_4)^2~~~~~t=(p_1-k)^2~~~~~u=(p_2-k)^2~~~~~~~q_T^2=ut/s
  \end{equation}
  and
  \begin{equation}
    z=M^2/s\, .
    \end{equation}
Since there is a lower limit on the ratio $r=q_T/M$ we can safely work in $d=4$ dimensions.
The differential cross reads
\begin{equation}
  \label{eq:diff}
   \frac{d^2 \sigmah_{q\bar q}}{d M^2 d q_T^2} = \frac{1}{32s^2}\frac{1}{(2\pi)^4}  \frac{1}{\sqrt{(1 -
   z)^2 - 4z q_T^2/M^2}} \sqrt{1 - \frac{4 m^2}{M^2}} \int d \Omega | \mathcal{M} |^2
\end{equation}
and the angular integral is defined in the centre-of-mass frame of the final-state leptons.
By integrating Eq.~(\ref{eq:diff}) over $q_T^2$ and $M^2$ and keeping into account the phase space constraints we obtain
\begin{equation}
  \frac{d \sigmah_{q\bar q}}{d r_{\text{cut}}^2} =-\frac{1}{32}\frac{1}{(2\pi)^4} \int_{\zmin}^{\zmax} \frac{z\, dz}{\sqrt{(1 -
   z)^2 - 4 z \rcut^2}} \sqrt{1 - \frac{\zmin}{z}} \int d \Omega | \mathcal{M} |^2 
  \label{dsdrcut2}.
\end{equation}
where
\begin{equation}\label{eq:zlimits}
  \zmin=\f{4m^2}{s}~~~~~~~~~~~~~\zmax=1-2\rcut\sqrt{1+\rcut^2}+2\rcut^2\, .
  \end{equation}
The matrix element squared $|{\cal M}|^2$ can be divided into three separate gauge invariant contributions: final state radiation, initial state radiation and interference.
The interference contribution is odd under the exchange $p_3\leftrightarrow p_4$ and therefore vanishes after angular integration.
We now discuss the final- and initial-state contributions in turn.

\subsection{Final-state radiation}
\label{sec:fin}

\noindent The integration of the matrix element squared corresponding to final state radiation over the angular variables can be carried out along the lines of Ref.~\cite{Beenakker:1988bq}.
After partial fractioning, the required angular integrals have the form
\begin{equation}
I^{ (k, l)} = \int_0^{\pi} \sin \vartheta_1 d \vartheta_1 \int_0^{\pi} d
   \vartheta_2 (a + b \cos \vartheta_1)^{- k} (A + B \cos \vartheta_1 + C \sin
   \vartheta_1 \cos \vartheta_2)^{- j}
\end{equation}
where the coefficients $a, b, A, B, C$ are functions of the invariants $s,M^2, u, t$.
The ensuing contribution to $d\sigmah_{q\bar q}/d\rcut^2$ can be expressed in the following form
\begin{equation}
\frac{d \sigmah^{\rm FS}_{q\bar q}}{d r_{\text{cut}}^2} = - \frac{4 \alpha^3e_q^2}{3s}\int_{\zmin}^{\zmax}dz \left[ \frac{K_1(z;\zmin)}{(1-z)^2\sqrt{(1 - z)^2 - 4 z r_{\text{cut}}^2}} + \frac{K_2(z;\zmin)r_{\text{cut}}^2}{(1-z)^4\sqrt{(1 - z)^2 - 4 z r_{\text{cut}}^2}} \right]
\label{eq:finstate}
\end{equation}
in terms of two coefficient functions, $K_1$ and $K_2$, which are regular at $z=1$ (soft limit) and do not depend on the cut-off parameter $r_{\text{cut}}$: 
\begin{equation}
\begin{split}
  K_1(z;\zmin) &= -\left[\zmin z^2+z(1+z)^2\right]\sqrt{1-\frac{\zmin}{z}} \\ & + z(1+z^2+\zmin z-\frac{\zmin^2}{2})\ln{\frac{\displaystyle 1+\sqrt{1-\frac{\zmin}{z}}}{\displaystyle 1-\sqrt{1-\frac{\zmin}{z}}}},
\end{split}
\end{equation}
and 
\begin{equation}
\begin{split}
  K_2 (z;\zmin) &= 2 z^2 \Bigg\{ \left [1 + z (6 + z) + \zmin z\right]\sqrt{1-\frac{\zmin}{z}} \\ & - \left( 1 + z^2 + \zmin (2 + z)  - \frac{\zmin^2}{2}\right) \ln{\frac{\displaystyle 1+\sqrt{1-\frac{\zmin}{z}}}{\displaystyle 1-\sqrt{1-\frac{\zmin}{z}}}}\Bigg\}.
\end{split}
\end{equation}
In the small-$\rcut$ limit the integral in Eq.~(\ref{eq:finstate}) can be computed by using the expansions
\begin{align}\label{eq:expansion_fsr}
  \f{\Theta(\zmax-z)\Theta(z-\zmin)}{(1-z)^2\sqrt{(1 - z)^2 - 4 z \rcut^2}}&=\f{1}{4}\delta(1-z)\f{1}{\rcut^2}+\f{\pi}{8}\left[\delta(1-z)+2\delta^\prime(1-z)\right]\f{1}{\rcut}+{\cal O}(1)\nn\\
      \f{\Theta(\zmax-z)\Theta(z-\zmin)\rcut^2}{(1-z)^4\sqrt{(1 - z)^2 - 4 z \rcut^2}}&=\f{1}{24}\delta(1-z)\f{1}{\rcut^2}+\f{\pi}{64}\left[3\delta(1-z)+2\delta^\prime(1-z)\right]\f{1}{\rcut}+{\cal O}(1)
\end{align}
and we obtain for the $\rcut$ dependence of the partonic cross section
\begin{equation}\label{eq:final_result_fsr}
\begin{split}
  \sigmah^{\rm FS}_{q\bar q}(s;r_\text{cut})&=\sigma_0(s)\f{\alpha}{2\pi}\bigg\{\left[2-\frac{(1+\beta^2)}{\beta}\ln\frac{1+\beta}{1-\beta}\right] \ln{(\rcut^2)}\\
  &- \frac{3\pi}{8} \left[ \frac{6(5-\beta^2)}{3-\beta^2} + \frac{-47+8\beta^2+3\beta^4}{\beta(3-\beta^2)}\ln\frac{1+\beta}{1-\beta} \right]r_\text{cut} \bigg \}+ O(r_\text{cut}^2)\\
&\equiv \sigmah^{\rm FS}_{\rm LP}(s;\rcut) + \sigmah^{\rm FS}_{\rm NLP}(s;\rcut) + O(r_\text{cut}^2)
\end{split}
\end{equation}
where we have dropped terms which do not depend on $\rcut$ and we have introduced the Born cross section
\begin{equation}
  \sigma_0(s)=\f{2\pi}{9s}\alpha^2 e_q^2\beta(3-\beta^2)
  \end{equation}
with $\beta=\sqrt{1-\f{4m^2}{s}}$.

Eq.~(\ref{eq:final_result_fsr}) shows that the final-state contribution to the NLO cross section, integrated down to $\rcut$, contains the expected single logarithmic term in $\rcut$, which is due to soft emission and will be cancelled by the subtraction
counterterm (more precisely, by the abelian limit of the term in the second line in Eq.~(\ref{eq:sig11})). The next-to-leading power contribution $\sigmah^{\rm FS}_{\rm NLP}(s;\rcut)$ is linear in $\rcut$ and it is responsible for the behavior observed in Figure~\ref{figs:rcut-var-EW-nocuts}.

\subsection{Initial-state radiation}
\label{sec:ini}

\noindent The integration of the matrix element squared corresponding to initial-state radiation over the angular variables is straightforward and we obtain
\begin{equation}
  \frac{d \sigmah^{\rm IS}_{q\bar q}}{d r_{\text{cut}}^2} = -\frac{4 \alpha^3e_q^4}{9s}\int_{\zmin}^{\zmax}dz \left[ \frac{K_3(z;\zmin)}{\rcut^2\sqrt{(1 - z)^2 - 4 z r_{\text{cut}}^2}} + \frac{K_4(z;\zmin)}{\sqrt{(1 - z)^2 - 4 z r_{\text{cut}}^2}} \right]
  \label{eq:isr}
\end{equation}
where the coefficient functions $K_3$ and $K_4$ now read
\begin{equation}
  K_3(z;\zmin)=\sqrt{1-\f{\zmin}{z}}\left(z+\f{\zmin}{2}\right)\f{1+z^2}{z^2}~~~~~~~~~~K_4(z;\zmin)=-2K_3(z;\zmin)\f{z}{1+z^2}\, .
\end{equation}
The coefficient function $K_3(z;\zmin)$ controls the most singular term, and is proportional to the Altarelli-Parisi splitting function.
In order to evaluate the integral in Eq.~(\ref{eq:isr}) we have to expand the distribution
\begin{equation}
  T(z,\rcut,\zmin)=\f{\Theta(z-\zmin)\Theta(\zmax-z)}{\sqrt{(1-z)^2-4z \rcut^2}} 
\end{equation}
in the small $\rcut$ limit. Since we know that linear terms in $\rcut$ are absent, we have to expand up to ${\cal O}(\rcut^2)$.
Such expansion is tricky, because the functions $K_3(z;\zmin)$ and $K_4(z;\zmin)$ contain a square root which vanishes at $z=\zmin$.
At variance with the case of final-state radiation,
this leads to spurious singularities when the distributions appearing in the expansion involve derivatives at $z=\zmin$.
We found it convenient to split the integration over $z$ as follows
\begin{equation}
  \int _{\zmin}^{\zmax} dz =  \int _{\zmin}^a dz + \int _a^{\zmax} dz, \quad \zmin < a < \zmax\, .
\end{equation}
The integral from $\zmin$ to $a$ can be safely computed by expanding in $\rcut$.
The integral from $a$ to $\zmax$ can be computed by using the expansion
\begin{align}
  T(z,\rcut,a)&=-\f{1}{2}\delta(1-z)\ln\rcut^2+\left(\f{1}{1-z}\right)_a+\ln(1-a)\delta(1-z)\nn\\
  &-\f{1}{2}\left(\delta^{(2)}(1-z)-2\delta^{(1)}(1-z)\right)\rcut^2\ln\rcut^2\nn\\
  &+\Bigg[\left(1+\ln{(1-a)}\right)\delta^{(2)}(1-z) - [1+2\ln{(1-a)}]\delta^{(1)}(1-z)- \frac{1}{2}\delta(1-z)\nn\\
    &+ \frac{1-2a}{(1-a)^2}\delta(z-a) - \frac{a}{1-a}\delta^{(1)}(z-a) + D^{(2)}(z,a) + 2 D^{(1)}(z,a)\Bigg]\rcut^2+{\cal O}(\rcut^4) .
  \label{eq:expansion}
  \end{align}
where we have defined the distributions $\delta^{(n)}(z-b)$, $\left(\frac{1}{1-z}\right)_a$, $D^{(1)}(z,a)$ and $D^{(2)}(z,a)$ through their action on a test function $f(z)$ as
\begin{align}
  \int_{0}^{1} dz f(z) \delta^{(n)}(z-b) &= (-1)^n f^{(n)}(b), \quad b\in[0,1],\\
  \int_{0}^{1} dz f(z) \left(\frac{1}{1-z}\right)_a  &= \int_a^1 dz \frac{f(z)-f(1)}{1-z},\\
  \int_{0}^{1} dz f(z) D^{(1)}(z,a) &= \int_a^1 dz \frac{f^{(1)}(z)-f^{(1)}(1)}{1-z},\\
  \int_{0}^{1} dz f(z) D^{(2)}(z,a) &= \int_a^1 dz \frac{zf^{(2)}(z)-f^{(2)}(1)}{1-z}\, .
\end{align}
By combining the two contributions $\zmin<z<a$ and $a<z<\zmax$ the dependence on $a$ cancels out and we obtain
for the $\rcut$ dependence of the partonic cross section
\begin{align}
\label{eq:final_result_isr}
~~~~\sigmah^{\rm IS}_{q\bar q}(s;r_\text{cut})&= \sigma_0(s)\frac{\alpha}{2\pi}e_q^2 \bigg\{ \ln^2{\rcut^2}- 4 \left( 2\ln{2} - \frac{4}{3} - \ln\frac{1-\beta^2}{\beta^2} - \frac{1}{\beta(3-\beta^2)}\ln\frac{1+\beta}{1-\beta} \right) \ln{\rcut^2}\nn\\ 
  & -\frac{3}{2}\frac{(1+\beta^2)(1-\beta^2)^2}{\beta^4(3-\beta^2)} \left(1-4\ln{2} +2\ln\frac{(1-\beta^2)\rcut}{\beta^2}\right) \rcut^2 \bigg\} +..........\nn\\
  &\equiv \sigmah^{\rm IS}_{\rm LP}(s;\rcut) + \sigmah^{\rm IS}_{\rm NLP}(s;\rcut) +..........
\end{align}
where we have dropped terms which do not depend on $\rcut$ and the dots stand for terms that vanish faster than $\rcut^2$ as $\rcut\to 0$.
At variance with Eq.~(\ref{eq:final_result_fsr}), Eq.~(\ref{eq:final_result_isr}) contains a double and a single logarithmic term in $\rcut$, which will be cancelled by the subtraction counterterm.
As expected, the next-to-leading power contribution $\sigmah^{\rm IS}_{\rm NLP}(s;\rcut)$ is quadratic in $\rcut$, modulo logarithmic enhancements.

As a byproduct of our calculation, we can reobtain the power suppressed terms in the case of the production of a vector boson of mass $M$.
To get rid of the decay it is enough to carry out the $m\to 0$ limit, while the constraint on the mass of the vector boson $M$ eliminates the integration over the $z$ variable.
By using the expansion in Eq.~(\ref{eq:expansion}) with $a=0$ the contributions from the functions $K_3(z;\zmin)$ and $K_4(z;\zmin)$ agree with the results in Eq.~(4.7) and (4.8) in Ref.~\cite{Cieri:2019tfv}.

\subsection{Numerical validation}
\label{sec:num}

In order to check the results presented in Secs.~\ref{sec:fin}, \ref{sec:ini} we have numerically implemented the exact real emission contribution to the cross section
and the expansions in Eqs.~(\ref{eq:final_result_fsr}) and (\ref{eq:final_result_isr}).

\begin{figure}
  \centering
  \includegraphics[width=0.48\textwidth]{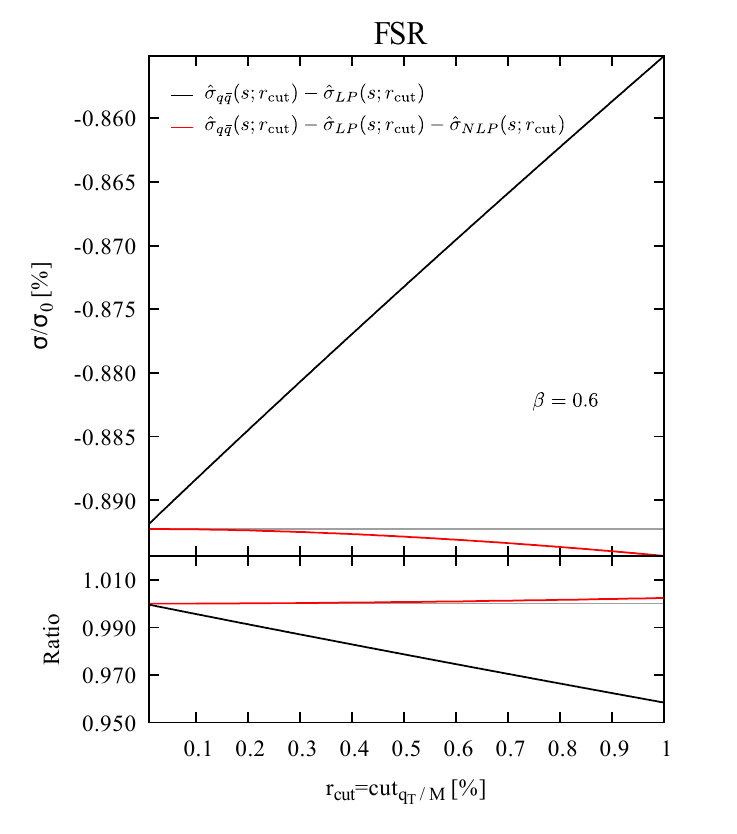}
  \includegraphics[width=0.48\textwidth]{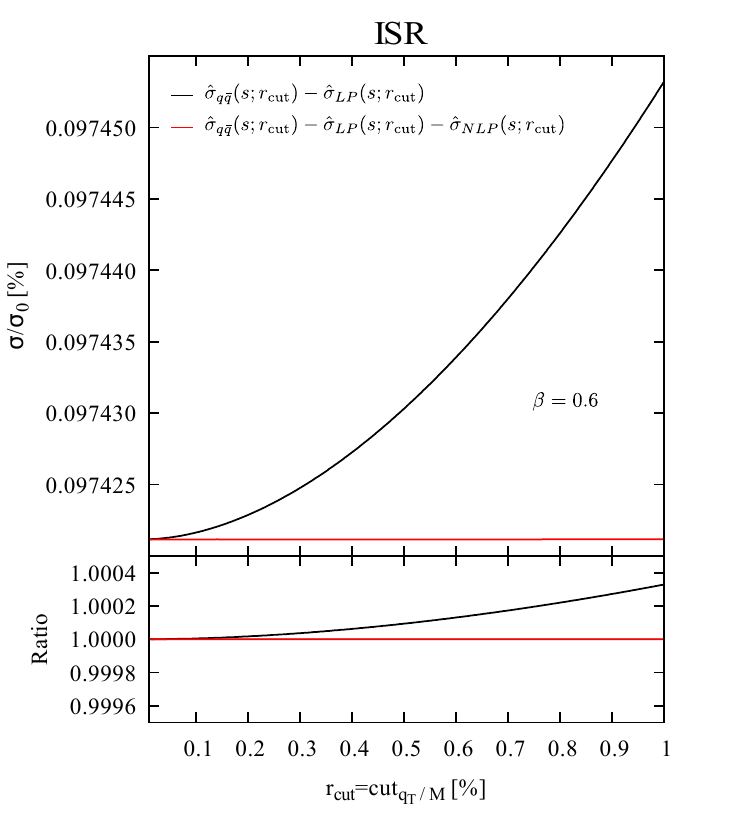}
  \caption{\label{figs:num-check:partonic-level} \small Subtracted partonic cross section for final-state radiation (left panel) and initial-state radiation (right panel). The solid lines represent the subtraction of the leading-power term, while the red solid line is obtained by subtracting also the next-to-leading power terms in Eq.~(\ref{eq:final_result_fsr}) and Eq.~(\ref{eq:final_result_isr}), respectively.
The upper panels show the result normalised to the Born cross section, while the lower panels show the result normalised to the $\rcut\to 0$ limit.
 The computation is carried out at fixed $\beta=0.6$. }
\end{figure}

In Figure~\ref{figs:num-check:partonic-level} we report the
exact real emission partonic cross section in the $q{\bar q}$ channel for $\beta=0.6$ as a function of $\rcut$ from which we have subtracted the leading-power contribution (black curve) and both the leading and next-to-leading power contributions (red curve). The numerical computation is separately carried out for the final-state radiation (left panel) and initial-state radiation (right panel) contributions.
Both for final-state radiation and initial-state radiation the leading-power contribution exactly matches the divergent behavior of the real emission cross section which is finite in the small-$\rcut$ limit.
The subtraction of the leading-power contribution exactly corresponds (up to quadratic terms in $\rcut$, see Eqs.~(\ref{eq:I1I2ap})) to the second term on the right hand side of Eq.~(\ref{eq:mainew}) and it is thus what is usually done in the standard $q_T$ subtraction procedure.
In the case of final-state radiation (left panel) the subtracted cross section exibits the expected linear behavior, while for initial-state radiation (right panel) the subtracted cross section scales quadratically with $\rcut$.
When besides the leading-power contribution, also the next-to-leading power (linear) term is subtracted the final-state subtracted cross section (red curve) behaves quadratically with $\rcut$, consistently with the result
in Eq.~(\ref{eq:final_result_fsr}). In the case of initial-state radiation, the additional subtraction of the next-to-leading power (quadratic) term makes the subtracted cross section almost independent on $\rcut$.

\subsection{Hadronic cross section}
\label{sec:had}

We now briefly comment upon the behavior of the hadronic cross section. In the following, we will show that when the fully inclusive cross section is considered, the convolution with the PDFs potentially introduces an additional linear term in $\rcut$.
In the case of final-state radiation such contribution could modify the parton level result. In the case of initial-state radiation such contribution could potentially change the power counting, by making the power correction linear. 
However, we will argue that, both for final-state and initial-state radiation, such additional term vanishes.

The real contribution to the hadronic cross section reads
\begin{equation}
  \label{eq:hadrcut}
  \sigma(S,\rcut) = \sum_{a,b}\int_0^1 dx_1 \int_0^1 dx_2f_a(x_1,\mu_F)f_b(x_2,\mu_F)\sigmah_{ab}(s,r_\text{cut})\delta(x_1x_2S-s)
\end{equation}
where $S$ is the hadronic CM-energy. The presence of a finite $\rcut$ implies that 
\begin{equation}
  s > \frac{4m^2}{\zmax}.
\end{equation}
where $\zmax$, defined in Eq.~(\ref{eq:zlimits}), behaves linearly with $\rcut$
\begin{equation}
\zmax=1-2\rcut+{\cal O}(\rcut^2)\, .
\end{equation}
The hadronic cross section in Eq.~(\ref{eq:hadrcut}) can be rewritten as
\begin{align}
  \label{eq:hadrcut2}
  \sigma(S,\rcut) &= \sum_{a,b}\int_0^1 dx_1 \int_0^1 dx_2f_a(x_1,\mu_F)f_b(x_2,\mu_F) \Theta\left(x_1x_2S-\frac{4m^2}{\zmax}\right)\hat{\sigma}_{ab}\left(s=x_1x_2S,\rcut\right)\nn\\ 
                  &= z_0 \sum_{a,b}\int_{z_0}^{\zmax} \frac{dz}{z^2} \int_{\ln{\sqrt{{z_0}/{z}}}}^{-\ln{\sqrt{{z_0}/{z}}}} dy f_a\left(\sqrt{\frac{z_0}{z}}e^y,\mu_F\right) f_b\left(\sqrt{\frac{z_0}{z}}e^{-y},\mu_F\right) \hat{\sigma}_{ab}\left(s=\frac{4m^2}{z},\rcut\right)\nn \\
                  &\equiv   \sum_{a,b}\int_{z_0}^{\zmax} dz\, {\cal L}_{ab}(z,z_0;\mu_F)\,{\hat \sigma}_{ab}\left(s=\frac{4m^2}{z},r_\text{cut}\right) 
\end{align}
where in the last step, we have performed the change of variables 
\begin{equation}
  x_1 = \sqrt{\frac{z_0}{z}}e^y,\qquad   x_2 = \sqrt{\frac{z_0}{z}}e^{-y},\qquad z_0 \equiv \frac{4m^2}{S}\, .
\end{equation}
The presence of $\zmax$ as an upper integration limit in Eq.~(\ref{eq:hadrcut2})
could potentially induce an additional linear term in $\rcut$ when the hadronic cross section is evaluated.
However, the partonic cross section vanishes at the kinematical limit $z=\zmax$
\begin{equation}
  {\hat \sigma}_{ab}\left(s=\frac{4m^2}{\zmax},r_\text{cut}\right)=0\, .
\end{equation}
This is a sufficient mathematical condition to prevent the appearance of a further linear term through integration.
We thus conclude that, as anticipated, in the case of final-state radiation the linear term in $\rcut$ is completely driven by the parton level result, while for initial-state radiation the convolution with PDFs will not produce
linear terms in $\rcut$.

\subsection{Final-state radiation at next-to-leading power: beyond inclusive observables}
\label{sec:diff}

In Sec.~\ref{sec:fin}, we have established by means of an analytical computation that in the case of final-state radiation off massive emitters
the leading power corrections are linear in $\rcut$ and we have explicitly evaluated the coefficient of the linear term.
The feasibility of our computation and the methods involved crucially rely on the fact we consider a sufficiently inclusive observable as the total cross section.
In the following we propose a way to promote the calculation of the final-state radiation contribution to differential level.

Our starting point is the expansion of the real contribution to the differential cross section in the soft limit.
According to the discussion in \ref{App}, the leading soft contribution allows us to obtain the leading-power term in $\rcut$, while the next-to-soft contribution will allow us to obtain the next-to-leading power.
By inspection of Eq.~(\ref{eq:expansion_fsr}) which involves the derivative of the $\delta$-distribution, we indeed expect that higher order terms in the soft expansion 
contribute to the next-to-leading power. 

In the following we propose a strategy to numerically prove the above result, which in turns provides a procedure to compute the next-to-leading power in a fully differential way.
From the soft contributions
we can construct a local counterterm which cancels the singularities of the real cross section but does not contribute to the next-to-leading power. 
Then the subtracted cross section is finite in four dimensions and can be integrated numerically in the unresolved region $r < \rcut$.
By construction, since the standard $q_T$ subtraction counterterm does not lead to linear contributions in $\rcut$ (see Eqs.~(\ref{eq:I1I2ap})),
the combination of the standard $q_T$ subtraction formula for $r>\rcut$ with such an additional subtracted term for $r < \rcut$
will be free of linear terms in $\rcut$.

To construct the additional counterterm we need a mapping which reabsorbs the radiation into a Born-like configuration.  
Among the available mappings at NLO, we choose the mapping proposed in Ref.~\cite{Buonocore:2017lry}. It is a massive FKS \cite{Frixione:1995ms} mapping dedicated to the case of the radiative emission off final state massive emitters and present some peculiar features:
\begin{itemize}
  \item the radiation is reabsorbed in such a way not to modify the partonic CM energy;
  \item the energy of the radiation (in the CM frame), which controls the way the soft limit 
        is approached, appears explicitly among the variables of the mapping. 
\end{itemize}
In the above mapping, we identify an emitter and a radiated parton. The radiation variables are given by the radiation energy fraction 
$\xi = 2 E_\text{rad}/\sqrt{s}$ ($s$ is the partonic CM energy), the cosine of the angle between the emitter and the radiated partons $y$
and an azimuthal angle $\phi$ (we refer to Ref.~\cite{Buonocore:2017lry} for more details). 
The phase space reads
\begin{equation}
  d\Phi_R = d\Phi_B \times J(\xi,y,\phi) d\xi dy d\phi
\end{equation}
where $d\Phi_B$ is the Born phase space element and the jacobian $J$ is given in Eq.~(49) of Ref.~\cite{Buonocore:2017lry}, and reduces to $J^{(0)}=s\,\xi/(4\pi)^3$ in the soft limit.
The local soft counterterm is defined as
\begin{equation}
  \label{eq:localsoftct}
  d \sigmah^{CT}_{S} = d\sigmah_{LO}(\Phi_B) \times \frac{e^2}{4\pi^3s}\frac{d\xi}{\xi}dy d\phi\left[\frac{s-2m^2}{(1-\beta y_\text{phy})(1+\beta y_\text{phy})} -\frac{m^2}{(1-\beta y_\text{phy})^2} - \frac{m^2}{(1+\beta y_\text{phy})^2}\right]
\end{equation} 
where $\beta = \sqrt{1-{4m^2}/{s}}$ and $y_\text{phy}$ is the cosine of the physical angle between the emitted photon and the leptons in the Born configuration (in practice we have either $y_\text{phy}=y$ or $y_\text{phy}=-y$ \cite{Buonocore:2017lry}).  
The next-to-leading power correction as function of $\rcut$ is obtained by subtracting the new soft counterterm from the real emission contribution in the unresolved region $r<\rcut$
\begin{equation} \label{eq:NLP_master_differential}
  d\sigmah^{\rm FS}_{\text{NLP}}(\rcut) = d\sigmah^{\rm FS}\Theta(r_\text{cut} M(\Phi_R) - q_T) - d\sigmah^{CT}_S\Theta( r_\text{cut} \sqrt{s} - q_T )
\end{equation}
where in the argument of the second theta function, we take the soft limit $M(\phi_R) \to M(\Phi_B)=\sqrt{s}$.

\begin{figure}[htb]
    \includegraphics[width=0.48\textwidth]{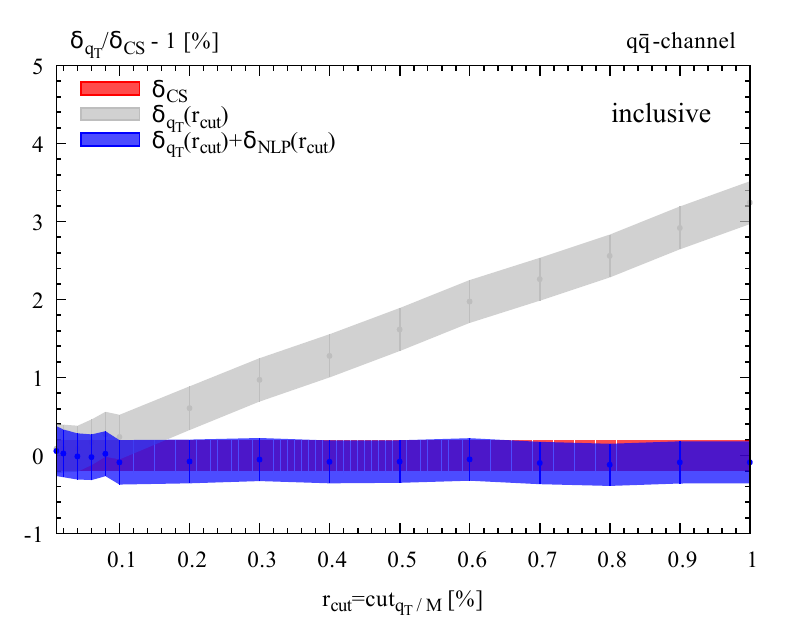}
    \includegraphics[width=0.48\textwidth]{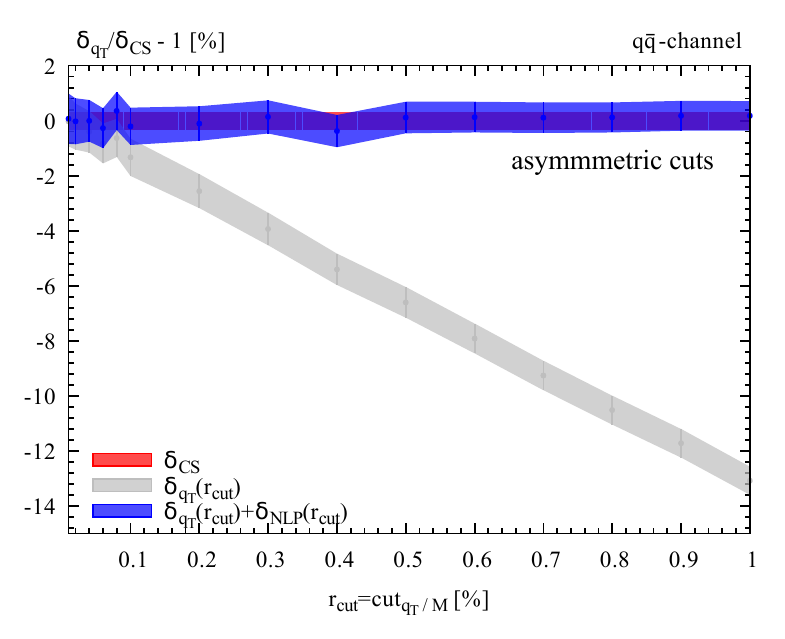}
    \caption{\label{fig:gamma+Z_lincorr}
      \small NLO EW correction as a function of $\rcut$ for the complete Drell-Yan process in the dominant $q\overline{q}$ diagonal channel without cuts (left panel) and with asymmetric cuts (right panel) at $7\,$TeV. The standard result obtained with $q_T$ subtraction (grey band) is compared with the result obtained by including the power suppressed contribution in Eq.~(\ref{eq:NLP_master_differential}).
      The NLO result is normalised to the $\rcut$-independent cross section computed with dipole subtraction.}
\end{figure}

The expression in Eq.~(\ref{eq:NLP_master_differential}) is fully differential, so that it can be used also in the case in which cuts on the final state are applied.
The contribution in Eq.~(\ref{eq:NLP_master_differential}) can be combined with the standard $q_T$ subtraction formula in Eq.~(\ref{eq:mainew}) to obtain an improved
subtraction procedure. 
In Figure~\ref{fig:gamma+Z_lincorr} we study the $\rcut$ dependence of the NLO EW correction to the complete Drell-Yan process when the $q_T$ subtraction formula is supplemented with
the next-to-leading power term in Eq.~(\ref{eq:NLP_master_differential}). We consider $pp$ collisions at $\sqrt{S}=7$ TeV and we compute the $\rcut$ dependent correction $\delta_{q_T}(\rcut)$ in the case
in which no cuts are applied (Figure~\ref{fig:gamma+Z_lincorr} (left)) and when asymmetric cuts on the transverse momenta and rapidities are applied: $p_{T,l^-}>25\,$GeV, $p_{T,l^+}>20\,$GeV and $|y_l|<2.5$ (Figure~\ref{fig:gamma+Z_lincorr} (right)).
We see that in both cases the linear dependence with $\rcut$ is nicely cancelled\footnote{As discussed in Sec.~\ref{sec:nloew}, when symmetric cuts are applied a linear dependence on $\rcut$ appears in the contribution from initial-state radiation.}.
The crucial point for this additional subtraction to be effective is that the additional counterterm in Eq.~(\ref{eq:localsoftct}) scales like $d\xi/\xi$, thereby leading to purely logarithmic contributions in $\rcut$. We have checked that alternative local subtractions which do not fulfill this property do not lead to a cancellation of the linear term.

We conclude this Section with few comments on the above results. The subtraction of the linear $\rcut$ behavior through Eq.~(\ref{eq:NLP_master_differential}) does not require any analytic integration.
It just requires an appropriate phase space mapping. The reader may of course argue that
there is no need to introduce the modification of Eq.~(\ref{eq:NLP_master_differential}) to achieve a smooth cancellation of the soft singularity.
Indeed, at NLO one can simply use a local subtraction scheme like FKS or dipole subtraction to carry out
the fully differential computation. Nonetheless, the strategy adopted here could prove itself useful when extending the computation to the mixed QCD-EW corrections with the $q_T$ subtraction formalism.
In this case, given that we aim at the computation of an effect of the order of few {\it per mille}, having a quadratic instead of linear $\rcut$ behavior
could dramatically improve the numerical control of the ${\cal O}(\alpha\as)$ contribution.

\section{Summary}
\label{sec:summa}

In this paper we have considered an application of the $q_T$ subtraction formalism to the production of massive lepton pairs in hadron collisions, and we have used this process as a case study to investigate the power suppressed contributions in the parameter $\rcut$. We have shown that $q_T$ subtraction
can be applied to evaluate NLO EW corrections to this process through a straightforward abelianisation procedure from heavy-quark production in QCD.
To obtain our numerical results and to discuss the $\rcut$ dependence we have focused on the case of a heavy lepton with mass $m_l=10$ GeV,
but the computation can be extended to lepton masses as small as the one of the muon without substantial complications, and we have been able to successfully reproduce results obtained with the numerical program {\sc SANC}.
Our calculation paves the way to possible applications to the computation of mixed QCD-EW corrections \cite{Dittmaier:2014qza,Dittmaier:2015rxo,deFlorian:2018wcj,Delto:2019ewv,Bonciani:2019nuy} and to NNLO QED corrections \cite{deFlorian:2018wcj} to the Drell-Yan process.

We have then studied the power suppressed contributions to the $q_T$ subtraction formula in the parameter $\rcut$. As is known, in the case of the inclusive cross section, initial state radiation leads to power corrections quadratic in $\rcut$,
and we have explicitly evaluated the corresponding NLO coefficient.
Generally speaking, linear power suppressed terms arise when cuts on the lepton transverse momenta are applied.
We have shown that, even in the case of the inclusive cross section, final state radiation leads to
a linear power correction in $\rcut$. We have explicitly computed the coefficient of such linear term in the case of the inclusive cross
section. By exploiting the purely soft nature of final state singularities,
we have also discussed how the result can be extended to differential distributions. The method used to carry out such extension could be used in future applications of the
$q_T$ subtraction formalism at ${\cal O}(\alpha\as)$.

\noindent {\bf Acknowledgements}. We are grateful to Stefano Catani for several helpful discussions and valuable comments.
This work was supported in part by the Swiss National Science Foundation (SNF) under contract 200020-169041, by MIUR under Project No.2015P5SBHT and by the INFN research initiative ENP.

\appendix

\gdef\thesection{Appendix \Alph{section}}
\section{Soft Power Counting}\label{App}
\gdef\thesection{\Alph{section}}

In this Appendix, we discuss the soft power counting for final-state radiation.
In the strictly soft limit, the phase space of the emitted photon with momentum $k$ exactly factorizes
\begin{equation}
  \label{eq:softpsp}
  d\Phi_3 = d\Phi_2 \times \frac{d^3 k}{(2\pi)^32k^0}.
\end{equation}
The leading power constribution to final-state radiation is given by the soft-factorisation formula
\begin{equation}
  |{\cal M}(p_1,p_2,p_3,p_4,k)|_{\rm FSR}^2\sim \left(e_3^2\,{\cal S}_{33}+e_4^2\,{\cal S}_{44}+2e_3 e_4\, {\cal S}_{34}\right) |{\cal M}(p_1,p_2,p_3,p_4)|^2
\end{equation}
where
\begin{equation}
  {\cal S}_{ij} = \frac{p_i \cdot p_j}{(p_i \cdot k)(p_j \cdot k)}.
\end{equation}
The power counting is more easily understood if we consider light-cone coordinates
\begin{equation}
  k^{\pm} = \frac{k^0 \pm k^3}{\sqrt{2}}~~~~~~~~~~d^4k= dk^+dk^-d^2\mathbf{k}_\perp
\end{equation} 
Then, the $1$-body phase space volume has the form
\begin{equation}
  \int \frac{d^4k}{(2\pi)^3} \delta_+(k^2) = \int \frac{dk^+dk^-d^2\mathbf{k}_\perp}{(2\pi)^3} \delta_+(2k^+k^- - \mathbf{k}_\perp^2) = \frac{1}{(2\pi)^3}\int_0^{\infty} \frac{dk^+}{2k^+} \int d^2\mathbf{k}_\perp 
\end{equation} 
with $k^-={\mathbf{k}_\perp^2}/{2k^+}$.
Considering for example the contribution from ${\cal S}_{34}$, the leading power unconstrained soft integral is given by
\begin{equation}
  \begin{split}
    I^{\text{soft}}_{34} &= \frac{1}{(2\pi)^3} \int_0^\infty \frac{dk^+}{2k^+} \int_0^\infty \frac{dk_\perp^2}{2} \int_0^{2\pi}d\theta\, {\cal S}_{34} \Theta(k^2_\perp - s r_\text{cut}^2)\\
    &= \frac{p_3 \cdot p_4}{(2\pi)^3} \int_{s r_\text{cut}^2}^\infty \frac{dk_\perp^2}{2} \int_0^\infty \frac{dk^+}{2k^+} \int_0^{2\pi} d\theta \frac{1}{p_\perp^2 k_\perp^2}\frac{1}{(a_3-\cos{\theta})(a_4+\cos{\theta})} \\
  \end{split}
\end{equation}
\begin{equation}
  a_i = \frac{1}{p_\perp k_\perp} \left( p_i^+\frac{k_\perp^2}{2k^+} + p_i^-k^+ \right ).
\end{equation}
In the above formula, we have enforced the soft kinematic with two back-to-back massive leptons. 
The azimuthal average is straightforward, after disentangling the product occuring in the denominator by means of the partial fractioning relation 
\begin{equation}
  \frac{1}{(a_3-\cos{\theta})(a_4+\cos{\theta})} = \frac{1}{a_3+a_4}\left( \frac{1}{a_3-\cos{\theta}} + \frac{1}{a_4+\cos{\theta}} \right),
\end{equation}
and it gives 
\begin{equation}\label{eq:pwcountI0}
    I^{\text{soft}}_{34} = \frac{p_3 \cdot p_4}{(2\pi)^2} \int_{s r_\text{cut}^2}^\infty \frac{dk_\perp^2}{2} \int_0^\infty \frac{dk^+}{2k^+} \frac{1}{p_\perp^2 k_\perp^2}\frac{1}{a_3+a_4}\sum_{i=3,4}\frac{1}{\sqrt{a_i^2-1}}.
\end{equation}
To make the scaling with the transverse momentum manifest, we apply the following change of variables at fixed $k_\perp$:
\begin{equation}
  x = \left(\f{k^+}{k_\perp}\right)^2,\quad d k^+ = k_\perp \f{dx}{2\sqrt{x}}\, .
\end{equation}
The soft integral becomes
\begin{equation}
I^{\text{soft}}_{34} = \frac{p_3 \cdot p_4}{(2\pi)^2}\frac{1}{\sqrt{2s}} \int_{s r_\text{cut}^2}^\infty \frac{dk_\perp^2}{k_\perp^2} \int_0^\infty  \frac{dx}{1+2x}\sum_{i=3,4}\frac{1}{\sqrt{4(p_i^-)^2x^2 + 2 (m^2 -p_\perp^2)x + (p_i^+)^2} }
\end{equation}
where $s$ is the partonic CM energy.\\
We can complete the calculation of the leading power contribution by performing the integration over the $x$ variable. The relevant integrals are of the form
\begin{equation}
  T(a,b,c) = \int_0^\infty \frac{dx}{1+2x}\frac{1}{\sqrt{a x^2 + 2 b x + c} } = \frac{1}{\sqrt{a-4b+4c}}\ln{\left[\frac{-2b+4c+2\sqrt{c}\sqrt{a-4b+4c}}{-a+2b+\sqrt{a}\sqrt{a-4b+4c}}\right]}
\end{equation}
under the conditions $b^2-ac<0$ and $a,c>0$. Then, it is straightforward to compute
\begin{equation}
   \int_0^\infty \frac{dx}{1+2x}\sum_{i=3,4}\frac{1}{\sqrt{4(p_i^-)^2x^2 + 2 (m^2 -p_\perp^2)x + (p_i^+)^2} } = \frac{1}{\sqrt{2}p}\ln{\frac{1+\beta}{1-\beta}}, \quad p=\sqrt{E^2-m^2}.
\end{equation}
We get the final expression
\begin{equation}\label{Isoft_rcut}
  I^\text{soft}_{34} = \frac{1}{4(2\pi)^2} \frac{1+\beta^2}{\beta}\ln{\frac{1+\beta}{1-\beta}} \int_{s r_\text{cut}^2}^\infty \frac{dk_\perp^2}{k_\perp^2}
\end{equation}
which exactly matches the coefficient of the leading logarithmic divergence proportional to the charge product $e_3e_4=-1$ in Eq.~\eqref{eq:final_result_fsr}.
The contributions from $I^{\rm soft}_{33}$ and $I^{\rm soft}_{44}$ can be obtained in a similar way and reproduce the remaining term in Eq.~\eqref{eq:final_result_fsr}.
The power counting for the linear power correction follows now easily observing that the energy of the radiation scales with the transverse momentum
\begin{equation}
  k^0 = \f{k^+ + k^-}{\sqrt{2}} = k_\perp \sqrt{\f{x}{2}}\left(1+\f{1}{2x}\right).
\end{equation}
This implies that corrections to the soft approximation will produce linear terms in $\rcut$.

\bibliography{biblioqt}
\end{document}